\documentclass[10pt,conference]{IEEEtran}
\usepackage{siunitx}
\usepackage{amsfonts,amsmath,amsthm}
\usepackage{graphicx,xspace,epsfig,syntonly,dsfont}
\usepackage{url,cite,bm,bbm}
\usepackage{algorithmicx}
\usepackage{balance}
\usepackage{subfigure}
\usepackage{stfloats}
\usepackage{diagbox}
\usepackage{multirow}
\newtheorem{remark}{Remark}

\usepackage{algorithm}
\usepackage{algpseudocode}

\long\def\symbolfootnote[#1]#2{\begingroup
\def\thefootnote{\fnsymbol{footnote}}
\footnote[#1]{#2}\endgroup}
\DeclareSIUnit \voltampere {MVA} %apparent power 
\IEEEoverridecommandlockouts
\allowdisplaybreaks[4]

\title{\Large \bf Unsupervised Deep Learning for AC Optimal Power Flow via Lagrangian Duality}
\author{
    \IEEEauthorblockN{Kejun Chen, Shourya Bose, and Yu Zhang} 
    \IEEEauthorblockA{Department of Electrical and Computer Engineering}
    \IEEEauthorblockA{University of California, Santa Cruz}
    \IEEEauthorblockA{Emails:\texttt\{kchen158,\, shbose,\,zhangy\}@ucsc.edu}
}

\IEEEoverridecommandlockouts\IEEEpubid{\makebox[\columnwidth]{978-1-6654-3540-6/22~\copyright~2022 IEEE \hfill} \hspace{\columnsep}\makebox[\columnwidth]{ }}

\begin{document}
\maketitle

\begin{abstract}
Non-convex AC optimal power flow (AC-OPF) is a fundamental optimization problem in power system analysis. The computational complexity of conventional solvers is typically high and not suitable for large-scale networks in real-time operation. Hence, deep learning based approaches have gained intensive attention to conduct the time-consuming training process offline. Supervised learning methods may yield a feasible AC-OPF solution with a small optimality gap. However, they often need conventional solvers to generate the training dataset. This paper proposes an end-to-end unsupervised learning based framework for AC-OPF. We develop a deep neural network to output a partial set of decision variables while the remaining variables are recovered by solving AC power flow equations. The fast decoupled power flow solver is adopted to further reduce the computational time. In addition, we propose using a modified augmented Lagrangian function as the training loss. The multipliers are adjusted dynamically based on the degree of constraint violation. Extensive numerical test results corroborate the advantages of our proposed approach over some existing methods. 
\end{abstract}
%---------------------------
\section{Introduction}\label{sect:intro}
AC optimal power flow (AC-OPF) is a fundamental problem for efficient and reliable operation and planning in electric power networks. AC-OPF minimizes an objective function (e.g., the total generation cost) subject to operational constraints, including nodal power balance and branch flow equations as well as inequality constraints regarding limits of power generations, voltage phasors, and branch flows. AC-OPF problems are typically non-convex due to the highly non-linear power balance equations.

Various approaches have been proposed to solve AC-OPF problems, e.g., convex relaxation and approximation methods. Inexact convex relaxations provably yield infeasible solutions \cite{Exactness}. Approximation methods such as DC-OPF models typically linearize the AC power flow (AC-PF) equations \cite{linear_opf}. The optimization problems with those approximate models can be solved rapidly for large-scale systems, but obtaining AC feasible solutions is difficult \cite{DC}. Conventional optimization solvers (e.g., Matpower interior point solver (MIPS) \cite{matpower}) may provide an AC-feasible solution. But they are generally not scalable for real-time operations. 

Recently, supervised or unsupervised deep learning based approaches have been proposed as AC-OPF solvers. The main motivation is to quickly yield a high-quality solution for real-time system operation by shifting the heavy computational burden to the offline training phase. Consider an OPF problem, the mapping from its input (active and reactive power demand) to an optimal solution as the output (power generation, voltage phasors, etc) is a very complicated function. Supervised learning methods estimate such a function based on the available input-output training data points. Existing supervised learning techniques can be classified into hybrid and stand-alone approaches.

Hybrid approaches focus on improving the performance of conventional solvers with the help of DNNs. For example, we can classify active/inactive constraints and reduce the problem size by removing the inactive ones \cite{meta}. Some algorithms provide a warm-start initial point for conventional solvers \cite{Smart-PGSim}.
In contrast, stand-alone approaches employ end-to-end deep learning frameworks that can directly output an optimal solution. Some DNN methods obtain all decision variables simultaneously while ignoring power balance equations, which may lead to load mismatch \cite{Owerko} and \cite{Fioretto2020}. Other methods first output a partial set of decision variables via a DNN, then obtain the remaining variables by dealing with the equality constraints. For example, \cite{DeepOPF-V} and \cite{Rahman} predict voltage phasors, and then compute active and reactive power generations using AC-PF equations. However, in this way the power balance at load buses may not be satisfied. \cite{Pan2020DeepOPFAF} predicts active power generations and voltage magnitudes of the generator buses and the reference bus. The remaining decision variables are recovered by solving the AC-PF equations, which guarantees the nodal power balance.

It is worth noting that supervised learning based methods need conventional solvers to build large training datasets, which takes extra time and may have suboptimal solutions. To bypass this limitation, there is an increasing interest in unsupervised learning frameworks without the aid of conventional solvers. Unsupervised learning methods can also incorporate variable splitting; e.g., \textit{NGT}~\cite{huang2021} and \textit{DC3}~\cite{donti2021dc3}. These two methods leverage multi-task learning by using a joint training loss function. However, the challenge is that increasing the weight of one task may deteriorate the performance of the others. Therefore, tuning the weighting parameters plays a critical role in finding a good tradeoff among all tasks. 

Inspired by previous works, we develop an end-to-end unsupervised learning framework with variable splitting. 
The main contribution of our paper is two-fold:
\begin{itemize}
    \item We propose to use a modified augmented Lagrangian function as the training loss, which contains the generation cost and penalty terms for constraint violation. The penalties involve Lagrangian multipliers, which serve a role of the weighting parameters. The multipliers are dynamically adjusted according to the degree of constraint violations during the training process. 
    \item  We adopt the fast decoupled power flow (FDPF) solver \cite{FDPF} in the framework. The proposed method can significantly speed up the computational time compared with conventional solvers, which is appealing for many real-time operations. 
\end{itemize}

\section{Problem Formulation} \label{sec:pf}
In this section, we formulate the AC-OPF problem and rewrite it as an optimization problem with inequality constraints only. In addition, we show how to use augmented Lagrangian relaxation to solve the reformulated problem.

%-------------
\subsection{AC-OPF Problem Formulation}
Consider a power network consisting of $N$ buses (denoted by set $\mathcal{N}$) and $M$ transmission lines (denoted by set $\mathcal{M}$). There are three different types of buses: the set of $N_d$ load buses denoted by $\mathcal{N}_d$, the set of $N_g$ generator buses denoted by $\mathcal{N}_g$, and one reference bus. As shown blow, the AC-OPF aims at minimizing the total generation cost while satisfying a set of operational constraints \cite{Cain12}.
\begin{subequations}
\label{AC-PF}
\begin{align}
    \min_{\mathbf{V}, \boldsymbol{\theta}, \mathbf{P}_g, \mathbf{Q}_g} \quad & \sum_{i} c_i(P_{g, i}) \label{oj}\\
    \textrm{s.t.} \quad & P_{g, i} - P_{d, i} = V_i \sum_{j=1}^N V_j (G_{ij}\cos \theta_{ij} + B_{ij}\sin \theta_{ij} )  \label{pf1} \\ 
    & Q_{g, i} - Q_{d, i} = V_i \sum_{j=1}^N V_j (G_{ij}\sin \theta_{ij} - B_{ij}\cos \theta_{ij} )  \label{pf2}\\ 
    & P_{ij} = -G_{ij}V_i^2 + V_iV_j(G_{ij}\cos \theta_{ij} + B_{ij}\sin \theta_{ij})  \label{bf1}\\
    & Q_{ij} = B_{ij}V_i^2 + V_iV_j(G_{ij}\sin \theta_{ij} - B_{ij}\cos \theta_{ij})  \label{bf2} \\
    & P_{ij}^2 + Q_{ij}^2 = |S_{ij}|^2,\, \forall  (i, j)\in \mathcal{M}  \label{bf3} \\
    & |S_{ij}|^2 \leq (S_{ij}^{\text{max}})^2,\, \forall  (i, j)\in \mathcal{M}  \label{bf4} \\
    & P_{g, i}^{\text{min}} \leq P_{g, i} \leq P_{g, i}^{\text{max}},\, \forall  i\in \mathcal{N} \setminus \mathcal{N}_d   \label{pg}\\
    & Q_{g, i}^{\text{min}} \leq Q_{g, i} \leq Q_{g, i}^{\text{max}},\, \forall  i\in \mathcal{N} \setminus \mathcal{N}_d  \label{qg}\\
    & V_{i}^{\text{min}} \leq V_{i} \leq V_{i}^{\text{max}},\, \forall  i\in \mathcal{N} \label{v} \\
    & \theta_{\text{ref}} = 0  \label{ang} \\
    & P_{g, i}=Q_{g, i} = 0,\, \forall  i \in \mathcal{N}_d.  \label{no_gen}
\end{align} 
\end{subequations}
The objective function \eqref{oj} is the total active power generation cost, where $c_i(\cdot)$ is the generation cost of unit $i$. $P_{g, i}$, $Q_{g, i}$, $P_{d, i}$ and $Q_{d, i}$ denote the active and reactive power generations and load demands at bus $i$. $V_i$ is the voltage magnitude of bus $i$. $\theta_{ij} := \theta_i - \theta_j$ is the voltage angle difference between bus $i$ and $j$. $P_{ij}$ and $Q_{ij}$ denote the active and reactive branch flows from bus $i$ to bus $j$. $G_{ij}$ and $B_{ij}$ are the real and imaginary parts of the $(i,j)$-th element of the nodal admittance matrix $\mathbf{Y} \in \mathbb{C}^{N\times N}$, respectively. Equality constraints \eqref{pf1} and \eqref{pf2} are nodal power balance equations. \eqref{bf1} and \eqref{bf2} represent $2M$ branch flow balance equations. \eqref{bf3} and \eqref{bf4} depict the squared apparent power flow and its upper bound. In addition, \eqref{pg}--\eqref{v} are the box constraints of active/reactive power generations and voltage magnitudes. The voltage angle of the reference bus is set to zero in \eqref{ang}. Finally, \eqref{no_gen} indicates that load buses have no power generation. 

\subsection{Variable Splitting}

Let $\mathbf{x} = [(\mathbf{P}_d)_{\mathcal{N}}; (\mathbf{Q}_d)_{\mathcal{N}}] \in \mathbb{R}^{2N}$ collect the load demands of all buses. Let $\mathbf{y} = [(\mathbf{P}_g)_{\mathcal{N}_g}; (\mathbf{V})_{\mathcal{N}_g}; V_{\text{ref}}; \theta_{\text{ref}}] \in \mathbb{R}^{2N_g + 2}$, where $(\mathbf{P}_g)_{\mathcal{N}_g}$ and $(\mathbf{V})_{\mathcal{N}_g}$ are the active power generations and voltage magnitudes of generator buses. Finally, the remaining decision variables are denoted by $\mathbf{z}_1 = [(\mathbf{V})_{\mathcal{N}_d}; \boldsymbol{\theta}_{\mathcal{N}_g \cup \mathcal{N}_d}] \in \mathbb{R}^{2N_d + N_g}$ and $\mathbf{z}_2 = [P_{g, \text{ref}}; Q_{g, \text{ref}}; (\mathbf{Q}_g)_{\mathcal{N}_g}, \mathbf{S}_{ij}^2] \in \mathbb{R}^{N_g + M + 2}$, where $P_{g, \text{ref}}$ and $Q_{g, \text{ref}}$ are the active and reactive power generations of the reference bus, respectively. 

\begin{figure*}[t]
    \centering
    \includegraphics[scale=0.35]{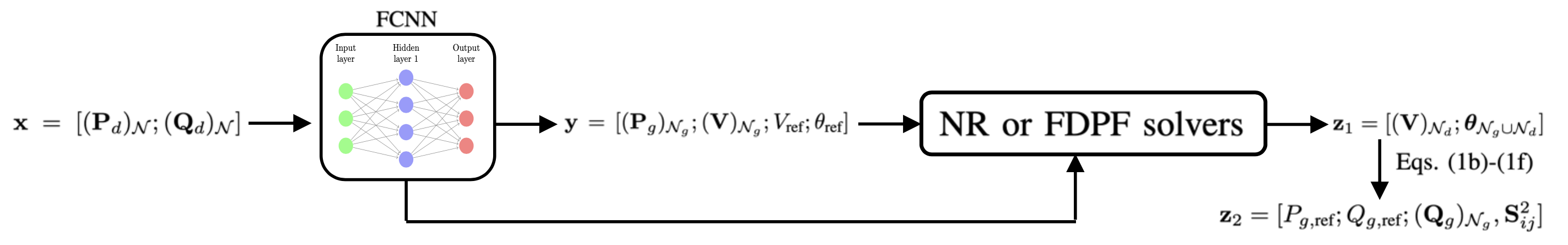}% second figure itself
    \caption{The proposed framework of an unsupervised deep learning model for solving AC-OPF.}
    \label{opf_nn3}
\end{figure*}

We develop a fully connected neural network (FCNN) to approximate the mapping from the input $\mathbf{x}$ to the partial decision variables $\mathbf{y}$. Once $\mathbf{x}$ and $\mathbf{y}$ are obtained, we can build $2N_d+N_g$ nodal power balance equations that are a subset of \eqref{pf1}--\eqref{pf2}. Therefore, $\mathbf{z}_1$ consisting of unknown voltage magnitudes and angles can be recovered by solving the equations via Newton-Raphson (NR) \cite{NR} or FDPF solvers. Once voltage magnitudes and angles of all buses become available, $\mathbf{z}_2$ can be uniquely determined by evaluating the equality constraints \eqref{pf1}--\eqref{bf3}. Clearly, splitting the decision variables in this way guarantees that the equality constraints in \eqref{AC-PF} are always satisfied. Let $\mathbf{u}(\cdot)$ denote the mapping from $\mathbf{y}$ to $\mathbf{z}_1$ and $\mathbf{z}_2$. The schematic of our proposed framework is shown in Fig. \ref{opf_nn3}. 

\subsection{Augmented Lagrangian Relaxation}
We rewrite the original AC-OPF problem \eqref{AC-PF} as a generic optimization problem with inequality constraints as follows:
\begin{subequations}
\label{gen}
\begin{align}
    \min \quad & f(\mathbf{y}, \mathbf{z}_2)  \label{oj_gen}\\
    \textrm{s.t.} \quad & \mathbf{h}(\mathbf{y}, \mathbf{z}_1, \mathbf{z}_2) \leq \mathbf{0},  \label{cons_gen} 
\end{align}
\end{subequations}
where $f(\mathbf{y}, \mathbf{z}_2)$ is the objective \eqref{oj}, and vector-valued function $\mathbf{h}(\cdot)$ include all inequality constraints \eqref{bf4}-\eqref{v}.

By introducing slack variables $\mathbf{s} \in \mathbb{R}^{M+2N+4(N_g+1)}$, we can convert the inequality constraints \eqref{cons_gen} to equality constraints:
\begin{align}
    \mathbf{h}(\mathbf{y}, \mathbf{z}_1, \mathbf{z}_2) + \mathbf{s} \odot \mathbf{s} = \mathbf{0}    
\end{align}
Based on the augmented Lagrangian function of equality constraints, \cite{Liu2021ANA} shows the augmented Lagrangian method solves the following unconstrained objective after eliminating $\mathbf{s}$:
\begin{multline}
    \label{augmented Lagrangian function}
    L(\mathbf{y}, \mathbf{z}_1, \mathbf{z}_2, \boldsymbol{\mu}) = f(\mathbf{y}, \mathbf{z}_2) \, +  \\
    \frac{1}{2\alpha} \mathbf{1}^\top  \big((\text{ReLu}(\boldsymbol{\mu} +  \alpha \mathbf{h}(\mathbf{y}, \mathbf{z}_1, \mathbf{z}_2)))^2 - \boldsymbol{\mu} \odot \boldsymbol{\mu}\big) \, ,
\end{multline}
where $\boldsymbol{\mu} \in \mathbb{R}^{M+2N+4(N_g+1)}$ collects Lagrangian multipliers associated with all inequality constraints; $\text{ReLu}(\cdot)$ is the element-wise rectified linear unit; $\odot$ denotes the Hadamard product; $\mathbf{1}$ is the all-one column vector with the same length of $\boldsymbol{\mu}$; and $\alpha$ is a positive constant coefficient. 

The dual function of \eqref{augmented Lagrangian function} is given by minimizing the Lagrangian function with respect to the primal variables:
\begin{equation}
    \label{dual function} 
    g(\boldsymbol{\mu}) = \min_{\mathbf{y}, \mathbf{z}_1, \mathbf{z}_2} L(\mathbf{y}, \mathbf{z}_1, \mathbf{z}_2, \boldsymbol{\mu}). 
\end{equation}
The dual problem maximizes the dual objective in order to find the best lower bound of $f(\mathbf{y}, \mathbf{z}_2)$ as shown below:
\begin{align}
    \max_{\boldsymbol{\mu} \geq \mathbf{0}} \quad  g(\boldsymbol{\mu}). 
    \label{dual problem}
\end{align}
The dual problem can be solved by various Lagrangian methods. Consider the primal-dual approach which updates primal and dual variables sequentially at each iteration. Given $\boldsymbol{\mu}_k$ as the multiplier vector at the $k$-th iteration, the primal update solves the unconstrained problem \eqref{dual function} to obtain primal variables $\{\mathbf{y}_k, \mathbf{z}_{1, k}, \mathbf{z}_{2, k}\}$. Then, the multipliers can be updated via projected (sub-)gradient ascent fashion: 
\begin{align}
    \boldsymbol{\mu}_{k+1} &= \text{ReLu}\big(\boldsymbol{\mu}_{k} + \alpha \mathbf{h}(\mathbf{y}_k, \mathbf{z}_{1, k}, \mathbf{z}_{2, k})\big).
        \label{dual update}
\end{align}

% %-----------------------
\section{Proposed Approach}\label{sec:approach}
This section presents the unsupervised deep learning framework designed for solving the AC-OPF problem. Moreover, we modify the augmented Lagrangian function to serve as the training loss function of the FCNN. 

\subsection{Deep Learning Framework for Solving AC-OPF}
Given $\mathbf{x}$ and $\mathbf{y}$, the proposed framework can obtain $\mathbf{z}_1$ and $\mathbf{z}_2$ that satisfy all equality constraints. Therefore, the critical step is to approximate the mapping from the input load demands $\mathbf{x}$ to the partial decision variables $\mathbf{y}$. FCNNs are  composed of a sequence of linear layers and activation functions, which can approximate any function theoretically (cf. the universal approximation theorem \cite{Hornik1989}). Therefore, we consider constructing an FCNN to approximate the complicated mapping $\mathbf{y} =\mathcal{O}_\mathbf{W}(\mathbf{x})$, where $\mathbf{W}$ represents the weights of the FCNN.

For an FCNN that has one hidden layer, the mapping between the input $\mathbf{x}$ and the output $\mathbf{y}$ can be expressed as:
\begin{align}
    \mathbf{y} = \mathcal{B}\big(\text{Sigmoid}(\mathbf{W}_2\text{ReLu}(\mathbf{W}_1\mathbf{x}))\big) \, ,
\end{align}
where $\mathbf{W}_i$ is the weight matrix of $i$-th linear layer. $\text{Sigmoid}(\cdot)$ is chosen to be the activation function of the output layer. $\mathcal{B}(\cdot)$ is a simple linear operator ensuring that the output variables satisfy their box constraints. For example, let $\beta \in [0,1]$ denote the FCNN's output for voltage magnitude $V_i$. Then,  $V_i \in [V_{i}^{\text{min}}, V_{i}^{\text{max}}]$ can be recovered via $\mathcal{B}(\cdot)$ as:
\begin{align}
    V_i= \mathcal{B}(\beta) := \beta V_{i}^{\text{min}} + (1-\beta) V_{i}^{\text{max}} \,.
\end{align}
We conduct similar transformations for all decision variables in $\mathbf{y}$ to satisfy the related box constraints.

\subsection{Combine Deep Learning with Lagrangian Duality}
Under our framework shown in Fig.~\ref{opf_nn3}, $\mathbf{y}$ is computed by the forward propagation, i.e., $\mathbf{y} := \mathcal{O}_\mathbf{W}(\mathbf{x})$. 
Now, plugging $\mathbf{y} = \mathcal{O}_\mathbf{W}(\mathbf{x})$ and $\{\mathbf{z}_1, \mathbf{z}_2\} = \mathbf{u}\left(\mathcal{O}_\mathbf{W}(\mathbf{x})\right)$ into the augmented Lagrangian \eqref{augmented Lagrangian function}, we  
obtain the Lagrangian function parameterized by the network weights $\mathbf{W}$ as 
\begin{equation}
    \label{dual function nn} 
     \mathcal{L}_{\mathbf{W}} := L(\mathcal{O}_\mathbf{W}(\mathbf{x}), \mathbf{u}(\mathcal{O}_\mathbf{W}(\mathbf{x})), \boldsymbol{\mu}).
\end{equation}
Combining DNN with the Lagrangian duality, this parameterized Lagrangian can be naturally served as the training loss function. That is, training the DNN by minimizing the loss function through backpropagation is equivalent to minimizing the parameterized Lagrangian over $\mathbf{W}$; see also \cite{Fioretto2020}:

To this end, our proposed scheme is listed as Algorithm \ref{alg}. Once the training process is completed, the trained FCNN can be employed to quickly predict $\mathbf{y}$ for any input $\mathbf{x}$ in the testing phase. The other decision variables $\mathbf{z}_1$ and $\mathbf{z}_2$ are determined by the equality constraints. 

\begin{remark}
It is worth emphasizing that the proposed method does not rely on the label $\mathbf{y}$, which is often obtained from conventional solvers. Hence, our approach belongs to the category of unsupervised learning. Based on the constraint violation degree (cf. \eqref{dual update}), the penalty term in the loss function is adjusted via the periodic update of the multiplier vector $\boldsymbol{\mu}$. Thanks to the Lagrangian duality theory, the proposed approach features a better training process than existing methods that adjust penalty weights heuristically.
\end{remark}

\begin{algorithm}[t]
\caption{Deep Learning Method via Lagrangian Duality}\label{alg}
\begin{algorithmic}[1]
\Require Dataset $\mathcal{X}$, coefficient $\alpha$, initial value of multiplier $\boldsymbol{\mu}_0$, maximum training epoch $n$, multiplier updating period $m$. \;
% \State $\boldsymbol{\mu}_0 \gets \mathbf{0}$. \;
\For{epoch $i = 1, 2, \ldots, n$}: \;
\State Sample data points $\mathbf{x} \in \mathcal{X}$. \;
\State Compute $\mathbf{y}$ through feedforward propagation. \;
\State Obtain $\mathbf{z}_1$ using NR or FDPF solver.
\State Compute $\mathbf{z}_2$ according to Eqs. \eqref{pf1}-\eqref{bf3}.
\State Calculate the loss \eqref{dual function nn}, and update $\mathbf{W}$ via backpropagation.
\State $\boldsymbol{\mu}_{i+1} \gets \boldsymbol{\mu}_{i}$. \;
\If {$i\mod m\equiv 0$}
\State {$\boldsymbol{\mu}_{i+1} \gets \text{ReLu}\big(\boldsymbol{\mu}_{i} + \alpha \mathbf{h}(\mathbf{y}, \mathbf{z}_1, \mathbf{z}_2)\big)$}. \;
\EndIf
\EndFor
\end{algorithmic}
\end{algorithm}
% %-----------------------
% %----------------------
\section{Numerical Results} \label{sec:tests}
We name the proposed methods as NR-Dual and FDPF-Dual, which use the NR and FDPF solvers, respectively. This section compares them with the conventional solver \textit{MIPS}, as well as two unsupervised learning based methods \textit{DC3} and \textit{NGT}. We test the IEEE-30 and IEEE-118 bus systems, which provide nominal values of load demands $(\tilde{\mathbf{P}}_d)_{\mathcal{N}}$ and $(\tilde{\mathbf{Q}}_d)_{\mathcal{N}}$. The load demand samples are uniformly distributed over $[0.9\tilde{\mathbf{P}}_{d}, 1.1\tilde{\mathbf{P}}_{d}]$ and $[0.9\tilde{\mathbf{Q}}_{d}, 1.1\tilde{\mathbf{Q}}_{d}]$, and collected in dataset $\mathcal{X}$.

%-------------------------
\subsection{Simulation Setup}
We generate 5,000 samples in dataset $\mathcal{X}$ with a training/validation/testing ratio of 10:1:1. The experiments are executed on a server with NVIDIA Titan RTX GPU with 25GB of RAM. The Adam optimizer is used to train the FCNN based on Pytorch 1.7.1. The maximum training epoch $n=1000$ and the mini-batch size is 32. The FCNN has one hidden layer, and the number of neurons of the hidden layer is 50 and 100 for the IEEE-30 and IEEE-118 bus systems, respectively. The power flow solvers stop iterations when the norm of load mismatches is less than $10^{-5}$. We update the Lagrangian multiplier every $10$ epochs instead of every epoch to help stabilize the training process \cite{Chatzos2020}. Finally, $\alpha=2$ is used in the simulations. 

\addtolength{\topmargin}{0.012in}

\subsection{Performance Criteria}
We evaluate the performance of our proposed method on the testing dataset based on four different metrics.
\begin{enumerate}
    \item \textbf{Optimality}: the total active power generation cost.
    \item \textbf{Feasibility}: the feasibility rate is calculated using the ratio of the number of satisfied inequality constraints to the total number of inequality constraints. Furthermore, we calculate the mean and maximum values of $\boldsymbol{\nu} := \text{ReLu}(\mathbf{h}(\mathbf{y}, \mathbf{z}_1, \mathbf{z}_2))$ to evaluate how much the constraints are violated. 
    \item \textbf{Load mismatch}: the relative error of the reconstructed load demands and the input load demands.
    \item \textbf{Computational efficiency}: the computational time. 
\end{enumerate}

\subsection{Test Results}
Our proposed work and the DC3 method use the second splitting framework to guarantee satisfying power balance equations. However, the NGT method uses the first splitting framework, which may lead to load mismatches at load buses. Besides, the FCNN's output dimension of the NGT method is $2N$, which is much greater than $2N_g+2$. Therefore, the FCNN's training using the NGT method takes a longer time.

\subsubsection{The feasibility performance of the proposed methods} 
\begin{figure*}[t]
    \centering
    \includegraphics[scale=0.55]{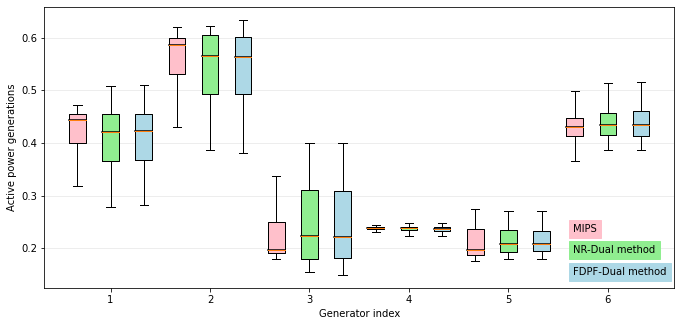}% second figure itself
    \caption{The boxplots of the active power generations on the IEEE-30 bus system.}
    \label{gene_boxplot}
\end{figure*}

Table \ref{nominal} shows the average nominal values of the decision variables in the per-unit systems (base value is \SI{100}{\voltampere}), which can serve as the reference to help evaluate the inequality constraints violations degree. As shown in Tables \ref{Feasibility evaluation 1} and \ref{Feasibility evaluation 2}, our proposed methods can obtain solutions whose mean values of violations degree are at least smaller by a fraction of $10^5$ than the nominal values. Similarly, the maximum values of violations degree are at least $10^3$ magnitude smaller than the nominal values. Besides, as shown in Fig.~\ref{gene_boxplot}, our proposed methods have similar generator allocation as MIPS. 

\begin{table}
\caption{The nominal values of decision variables}
\label{nominal}
\centering
\begin{tabular}{|c|c|c|}\hline
Test cases & Decision variables & Nominal values \\
\hline
\multirow{4}{*}{IEEE-30} & {$\mathbf{P}_g$} & 0.32 \\
\cline{2-3}

& {$\mathbf{Q}_g$} & 0.22  \\
\cline{2-3}

& {$\mathbf{V}$}  & 1.00 \\
\cline{2-3}

& {$\mathbf{S}_{ij}^2$} & 0.03  \\
\hline
\hline

\multirow{4}{*}{IEEE-118} & {$\mathbf{P}_g$} & 0.80  \\
\cline{2-3}

& {$\mathbf{Q}_g$} & 0.36 \\
\cline{2-3}

& {$\mathbf{V}$}  & 1.03  \\
\cline{2-3}

& {$\mathbf{S}_{ij}^2$} & 0.58 \\
\hline
\end{tabular}
\end{table}

\begin{table}
\caption{Feasibility evaluation of the NR-Dual method}
\label{Feasibility evaluation 1}
\centering
\begin{tabular}{|c|c|c|c|}\hline
Test cases & Decision variables & $\boldsymbol{\nu}$ Mean ($10^{-6}$) & $\boldsymbol{\nu}$ Max ($10^{-4}$) \\
\hline
\multirow{4}{*}{IEEE-30} & {$\mathbf{P}_g$} & 0 & 0\\
\cline{2-4}

& {$\mathbf{Q}_g$} & 0.89 & 0.05 \\
\cline{2-4}

& {$\mathbf{V}$}  & 0.08 & 0.02\\
\cline{2-4}

& {$\mathbf{S}_{ij}^2$} & 0.33 & 0.11 \\
\hline
\hline

\multirow{4}{*}{IEEE-118} & {$\mathbf{P}_g$} & 0 & 0 \\
\cline{2-4}

& {$\mathbf{Q}_g$} & 5.13 & 2.68\\
\cline{2-4}

& {$\mathbf{V}$}  & 0 & 0 \\
\cline{2-4}

& {$\mathbf{S}_{ij}^2$} & 2.77 & 4.30\\
\hline
\end{tabular}
\end{table}

\begin{table}
\caption{Feasibility evaluation of the FDPF-Dual method}
\label{Feasibility evaluation 2}
\centering
\begin{tabular}{|c|c|c|c|}\hline
Test cases & Decision variables & $\boldsymbol{\nu}$ Mean ($10^{-6}$) & $\boldsymbol{\nu}$ Max ($10^{-4}$) \\
\hline
\multirow{4}{*}{IEEE-30} & {$\mathbf{P}_g$} & 0 & 0 \\
\cline{2-4}

& {$\mathbf{Q}_g$} & 0 & 0  \\
\cline{2-4}

& {$\mathbf{V}$}  & 0 & 0\\
\cline{2-4}

& {$\mathbf{S}_{ij}^2$} & 0.53 & 0.19\\
\hline
\hline

\multirow{4}{*}{IEEE-118} & {$\mathbf{P}_g$} & 0 & 0\\
\cline{2-4}

& {$\mathbf{Q}_g$} & $3.37 $ & 1.68 \\
\cline{2-4}

& {$\mathbf{V}$}  & 0 & 0 \\
\cline{2-4}

& {$\mathbf{S}_{ij}^2$} & 3.10 & 4.50  \\
\hline
\end{tabular}
\end{table}

\subsubsection{Compare with the DC3 method \cite{donti2021dc3}}
As shown in equation \eqref{soft}, the loss function of the DC3 method consists of two parts, \emph{viz}., generation cost and penalty term for inequality constraints violations. The coefficient $\lambda$ serves as the weighting parameter to balance the relative importance of these two tasks. It is a predetermined hyperparameter and remains constant during the training process. Typically, a larger $\lambda$ promotes a solution with more minor inequality constraints violations but higher suboptimality in the generation cost. Therefore, we simulate the DC3 method with different values of $\lambda$ and show our proposed methods perform better comprehensively.
\begin{align}
\label{soft}
    \mathcal{L}_{\text{DC3}} := f(\mathbf{y}, \mathbf{z}_2) + \lambda \|\boldsymbol{\nu}\|_2^2 \, .
\end{align}

As shown in Table \ref{Performance}, for the IEEE-30 bus system, both NR-Dual and FDPF-Dual methods speed up the computation by 90x and  30x times than MIPS, respectively, which are at the expense of $0.15\%$ generation cost difference. DC3 with $\lambda = 1$ has the smallest generation cost. However, the mean and maximum values of $\boldsymbol{\nu}$ are 15 and 11 times greater than the proposed FDPF-Dual method. DC3 with $\lambda = 20$ yields a more infeasible and suboptimal solution compared with the proposed methods. Similarly, for the IEEE-118 bus system, the speedup factor is 68x and 220x times for NR-Dual and FDPF-Dual methods, respectively. The FDPF solver is three times faster than the NR solver in solving AC-PF equations. The feasibility rates of our proposed methods are both greater than $99\%$. In addition, the generation costs of the proposed approaches are only $0.19 \%$ and $0.16 \%$ greater than MIPS. 

\begin{table*}
\caption{Performance comparisons}
\label{Performance}
\centering
\begin{tabular}{|c|c|c|c|c|c|c|c|}\hline 
Test cases & Methods &  & Generation cost  & $\boldsymbol{\nu}$ Mean ($10^{-6}$) & $\boldsymbol{\nu}$ Max ($10^{-4}$) & Feasibility rate ($\%$) & Computational Times (s)  \\\hline
\multirow{10}{*}{IEEE-30} & \multirow{7}{*}{Different $\lambda$ in DC3} & 1 &  0.0645 & 2.78 & 2.09 & 99.44 & 0.13 \\  
\cline{3-8}
& & 2 & 0.0647  & 1.19  & 0.83 & 99.63 & 0.13 \\ 
\cline{3-8} 

& & 3 & 0.0648  & 0.47  & 0.55  & 99.85  & 0.13  \\ 
\cline{3-8} 

& & 5 & 0.0650 & 0.39  & 0.39 & 99.85  & 0.13\\ 
\cline{3-8} 

& & 10 & 0.0652  & 0.36  & 0.40 & 99.85 & 0.13  \\ 
\cline{3-8} 

& & 15 & 0.0655 & 0.28 & 0.27 & 99.85 & 0.13 \\ 
\cline{3-8} 

& & 20 & 0.0659 & 0.26 & 0.25 & 99.69  & 0.13  \\ 
\cline{2-8} 

& {\texttt{MIPS} optimizer} & - & 0.0646 & 0 & 0 & 99.99 & 12.05 \\ 
\cline{2-8}

& {NR-Dual method} & - & 0.0647 & 0.23 &  0.21 & 99.80 &  0.13 \\ 
\cline{2-8}

& {FDPF-Dual method} & - & 0.0647 & 0.18 &  0.19 & 99.78 & 0.39  \\ 
\hline 
\hline

\multirow{10}{*}{IEEE-118} & \multirow{5}{*}{Different $\lambda$ in DC3} & 1 & 13.145 & 23 & 72 & 97.66 & 0.52  \\  
\cline{3-8}

& & 3 & 13.156 & 8.20 & 31 & 98.48 & 0.52 \\ 
\cline{3-8} 

& & 5 & 13.161 & 7.92 & 32 & 98.62  & 0.52  \\ 
\cline{3-8} 

& & 10 & 13.174 & 4.27 & 18 & 98.94 & 0.52  \\ 
\cline{3-8} 

& & 15 & 13.181 & 2.88  & 13 & 99.16 & 0.52  \\ 
\cline{3-8} 

& & 20 & 13.184 & 2.96 & 13  & 99.11 & 0.52  \\ 
\cline{2-8} 

& {\texttt{MIPS} optimizer} & - & 13.137 & 0 & 0 & 99.95 & 35.33 \\ 
\cline{2-8}

& {NR-Dual method} & - & 13.162 & 1.66 & 9  & 99.21 & 0.52  \\ 
\cline{2-8}

& {FDPF-Dual method} & - & 13.158 & 1.45 & 8  & 99.17 & 0.16  \\ 
\hline 

\end{tabular}
\end{table*}

\subsubsection{Compare with the NGT method \cite{huang2021}}
The loss function of the NGT method consists of three parts: generation cost, penalty term of inequality constraint violations and load mismatch error. The FCNN's output consists of voltage magnitudes and angles of all buses. Using power balance equations, we can rebuild the active and reactive load demands at load buses, denoted by $\hat{\mathbf{x}}_d := [(\hat{\mathbf{P}}_d)_{\mathcal{N}_d}; (\hat{\mathbf{Q}}_d)_{\mathcal{N}_d}]$. Therefore, the error of load mismatch is given as: 
\begin{align}
    \mathcal{L}_d := \|\mathbf{x}_d - \hat{\mathbf{x}}_d\|^2_2 \, .
\end{align}
The training loss function of the NGT method is:
\begin{align}
    \mathcal{L}_{\text{NGT}}: = f(\mathbf{y}, \mathbf{z}_2) + \eta (1 - \tau) \|\boldsymbol{\nu}\|_2^2 + \eta \tau \mathcal{L}_d \, ,
\end{align}
where $\eta \in \mathbb{R}_{+}$ and $\tau \in [0, 1]$ are weighting parameters for balancing the three tasks.

In practice, a small relative error (typically less than $1\%$) of the load mismatch can be acceptable \cite{huang2021}. However, as shown in Tables \ref{Performance2} and \ref{Performance3}, the relative errors of load mismatch are much more significant than $1\%$ tested on the two benchmark systems. The computational times are 0.002s and 0.006s for the IEEE-30 and IEEE-118 bus systems, respectively.  

\begin{table*}
\caption{Performance of the NGT method on the IEEE-30 bus system}
\label{Performance2}
\centering
\begin{tabular}{|c|c|c|c|c|c|c|}\hline 
$\eta$ & $\tau$ & Generation cost & $\boldsymbol{\nu}$ Mean ($10^{-6}$) & $\boldsymbol{\nu}$ Max ($10^{-4}$) & Feasibility rate ($\%$) & Load mismatch ($\%$) \\\hline
\multirow{3}{*}{5} & 0.2 & 0.0642 & 0.23 & 0.03 & 99.99 & 5.29 \\  
\cline{2-7} 
& 0.5 & 0.0651 & 1.72 & 1.92 & 99.68 & 5.13  \\
\cline{2-7} 
& 0.8 & 0.0648 & 11.41 & 13.2 & 99.21 & 5.56 \\
\cline{1-7} 
\multirow{3}{*} {10} & 0.2 & 0.0646 & 0.20 & 0.25 & 99.95 & 5.55\\ 
\cline{2-7} 
 & 0.5 & 0.0662 & 1.14 & 1.41 & 99.77 & 5.52  \\ 
\cline{2-7} 
 & 0.8 & 0.0664 & 6.33 & 6.78 & 99.38 & 5.22  \\ 
\cline{1-7} 
\multirow{3}{*} {15} & 0.2 & 0.0654 & 0.46 & 0.57 & 99.88 & 5.50  \\ 
\cline{2-7} 
& 0.5 & 0.0665 & 0.88 & 0.99 & 99.80 & 5.22 \\ 
\cline{2-7} 
& 0.8 & 0.0670 & 4.29 & 4.81 & 99.48 & 5.08  \\ 
\hline
\end{tabular}
\end{table*}

\begin{table*}
\caption{Performance of the NGT method on the IEEE-118 bus system ($\tau = 0.5$)}
\label{Performance3}
\centering
\begin{tabular}{|c|c|c|c|c|c|}\hline 
$\eta$ & Generation cost & $\boldsymbol{\nu}$ Mean ($10^{-4}$) & $\boldsymbol{\nu}$ Max ($10^{-2}$) & Feasibility rate ($\%$) & Load mismatch ($\%$) \\\hline
5 & 8.80 & 2.88 & 2.56 & 79.10 & 140.40 \\  
\hline
10 & 12.71 & 0.43 & 0.25 & 99.20 & 21.93  \\ 
\hline
15 & 12.91 & 0.09 & 0.51 & 98.69 & 18.77 \\ 
\hline
20 & 13.07 & 0.00 & 0.20 & 99.50 & 16.94 \\ 
\hline
\end{tabular}
\end{table*}

%-+++
%--------------------------
\section{Conclusion}
This paper proposes a novel end-to-end unsupervised learning based framework to solve the challenging AC-OPF problem. Given load demands, the framework can quickly yield a high-quality feasible solution. Equality constraints are guaranteed to be satisfied via the decision variable splitting. We propose a modified augmented Lagrangian function as the training loss. The Lagrangian multipliers are updated periodically throughout the training process. Moreover, we incorporate the FDPF solver to further reduce the computational time of solving AC-PF equations. Extensive numerical results show that our proposed framework outperform state-of-the-art unsupervised learning based approaches. 

\nocite{*}
\bibliographystyle{IEEEtran}
\bibliography{LSrefs,IEEEabrv}

\end{document}